\documentclass[conference]{IEEEtran}
\IEEEoverridecommandlockouts

\usepackage[T1]{fontenc}
\usepackage{amsmath,amssymb,amsfonts}
\usepackage{graphicx}
\usepackage{textcomp}
\usepackage{relsize}

\usepackage{color}
\usepackage{pifont}
\usepackage{url}

\usepackage{tabularx}
\usepackage{booktabs}

\usepackage{adjustbox}
\usepackage{cite}
\usepackage{enumitem}

\usepackage{mathtools}
\usepackage{multirow}
\usepackage{comment}
\usepackage{booktabs}
\usepackage{algorithm}
\usepackage{subfigure}
\usepackage{algpseudocode}
\usepackage{listings}

\usepackage{placeins}

\usepackage{lscape}

\usepackage{comment}

\usepackage{pgfplots}
\pgfplotsset{compat=1.18}
\usepackage{pgfplotstable}
\usepackage{filecontents}
\usepackage{tikz}

\usetikzlibrary{patterns, shapes}

\usetikzlibrary{shapes.geometric, arrows.meta, positioning, fit, backgrounds, calc}

\def\BibTeX{{\rm B\kern-.05em{\sc i\kern-.025em b}\kern-.08em
    T\kern-.1667em\lower.7ex\hbox{E}\kern-.125emX}}

\definecolor{metric1}{RGB}{0,107,164}   % blue
\definecolor{metric2}{RGB}{255,128,14}  % orange
\definecolor{metric3}{RGB}{27,158,119}  % green
\definecolor{metric4}{RGB}{217,95,2}    % red-orange
\definecolor{metric5}{RGB}{117,112,179} % purple

\def\BibTeX{{\rm B\kern-.05em{\sc i\kern-.025em b}\kern-.08em
    T\kern-.1667em\lower.7ex\hbox{E}\kern-.125emX}}
\begin{document}

\title{A Practical Honeypot-Based Threat Intelligence Framework for Cyber Defence in the Cloud}

\author{\IEEEauthorblockN{Darren Malvern Chin}
 \IEEEauthorblockA{\textit{School of Information Technology,} \\
 \textit{Whitecliffe College,}\\
 Auckland, New Zealand \\
 20241314@mywhitecliffe.com}

 \and

 \IEEEauthorblockN{Bilal Isfaq}
 \IEEEauthorblockA{\textit{School of Information Technology,} \\
 \textit{Whitecliffe College,}\\
 Christchurch, New Zealand. \\
 bilali@whitecliffe.ac.nz}

 \and
 \IEEEauthorblockN{Simon Yusuf Enoch}
 \IEEEauthorblockA{\textit{School of Information Technology,} \\
 \textit{Whitecliffe College,}\\
 Wellington, New Zealand. \\
 0000-0002-0970-3621}
 }

\IEEEaftertitletext{\vspace{-1\baselineskip}}

\maketitle

% Later in the document:
\IEEEpubid{\raisebox{-0.7cm}{%
\begin{minipage}[t]{1.0\textwidth}
\raggedright
\large\textbf{\textcolor{blue!60!white}{2025 IEEE 35th International Telecommunication Networks and Applications Conference (ITNAC)}}
\end{minipage}}}
\IEEEpubidadjcol

\begin{abstract}
In cloud environments, conventional firewalls rely on predefined rules and manual configurations, limiting their ability to respond effectively to evolving or zero-day threats. As organizations increasingly adopt platforms such as Microsoft Azure, this static defense model exposes cloud assets to zero-day exploits, botnets, and advanced persistent threats.
In this paper, we introduce an automated defense framework that leverages medium- to high-interaction honeypot telemetry to dynamically update firewall rules in real time. The framework integrates deception sensors (Cowrie), Azure-native automation tools (Monitor, Sentinel, Logic Apps), and MITRE ATT\&CK-aligned detection within a closed-loop feedback mechanism. We developed a testbed to automatically observe adversary tactics, classify them using the MITRE ATT\&CK framework, and mitigate network-level threats automatically with minimal human intervention.

To assess the framework’s effectiveness, we define and applied a set of attack- and defense-oriented security metrics. Building on existing adaptive defense strategies, our solution extends automated capabilities into cloud-native environments. The experimental results show an average Mean Time to Block of 0.86 seconds—significantly faster than benchmark systems—while accurately classifying over 12{,}000 SSH attempts across multiple MITRE ATT\&CK tactics. These findings demonstrate that integrating deception telemetry with Azure-native automation reduces attacker dwell time, enhances SOC visibility, and provides a scalable, actionable defense model for modern cloud infrastructures.
\end{abstract}


\section{Introduction}

The growing adoption of cloud computing in enterprise environments has heightened the demand for security mechanisms that are both adaptive and intelligent. Dynamic firewall management and deception-based defense strategies are increasingly recognized as effective against threats such as zero-day exploits, polymorphic malware, and advanced persistent threats (APTs). However, traditional firewalls, which operate on static rules and require manual intervention, struggle to keep pace with evolving attacker tactics, particularly in dynamic networks (e.g., the cloud), where elasticity, scalability, and multi-tenancy, increase both complexity and the attack surface. Cyber deception (e.g., honeypots, decoy services, and fake credentials) offers a proactive approach, enabling cloud defenses to detect and thwart attacks before they can cause harm. In particular, medium to high-interaction honeypots can capture rich behavioral threat intelligence by recording attacker tactics in controlled environments~\cite{guan_honeyiot_2023,kandanaarachchi_honeyboost_2022,moric_advancing_2025}. This intelligence offers the potential for more adaptive and automated defense. However, most existing implementations treat honeypot data as post-incident forensic artifacts rather than live triggers for defensive action~\cite{amal_h-doctor_2023,fatima_enhancing_2024,dambrosio_smash_2025}. To effectively thwart attackers in real time, it is important to develop an approach that can automatically defend against them.

Several frameworks have attempted to address this problem. For example, H-DOCTOR~\cite{amal_h-doctor_2023} and SMASH SDN-MTD~\cite{dambrosio_smash_2025} combine deception telemetry with dynamic response, but are designed for on-premise or hybrid environments and lack cloud native orchestration, serverless automation, and tight integration with services such as Logic Apps, Sentinel, and Firewall Manager. Similarly, Tudosi et~al.~\cite{tudosi_design_2023} proposed a decentralized firewall with dynamic updates but without deception inputs or cloud-native compatibility. Other work, such as HoneyIoT~\cite{guan_honeyiot_2023} and HoneyBoost~\cite{kandanaarachchi_honeyboost_2022}, improves detection resilience but did not convert telemetry, such as command logs and source IPs, into automated firewall rules. As a result, most systems remain reactive rather than operating in real time.

Consequently, in this paper, we propose a cloud-native automation pipeline that leverages honeypot telemetry to generate actionable intelligence and enable real-time Azure firewall updates. Our framework converts raw attack data into predictive insights using native Azure tools, including Sentinel, enhanced with built-in AI/ML analytics for threat detection, Logic Apps for orchestration, and REST API integration for automated response. This closed-loop system enables rapid detection, enrichment, and automated enforcement of security policies, fully aligned with MITRE ATT\&CK tactics and techniques. By combining deception-based telemetry with intelligent automation, our approach overcomes the integration and responsiveness limitations of reactive systems. The contributions of this paper are as follows:
\begin{itemize}
    \item We design a framework that integrates medium to high-interaction honeypots with Azure services (Log Analytics, Sentinel, Logic Apps, Firewall Manager) for adaptive firewall automation, including MITRE ATT\&CK-aligned classification and enforcement triggers.
    \item We implement the framework in an Azure testbed with Cowrie honeypots, a closed-loop telemetry pipeline, Sentinel analytic rules, and Logic App workflows to apply firewall updates via NSG REST APIs.
    \item We evaluate the system using emulated and unsolicited attack conditions, measuring metrics such as Mean Time to Block (MTTB), attack success rates, and engagement time, validating its scalability and precision.
\end{itemize}

The rest of the paper is organized as follows: Section~\ref{sec:relatedwork} reviews background and related work; Sections~\ref{sec:proposed_approach} and~\ref{sec:evaluation} detail the proposed framework and evaluation metrics; Section~\ref{sec:Experiments} and Section~\ref{sec:MITRE} describe experiments, results, and analysis; and Section~\ref{sec:conclusion} concludes the paper.

\section{Related Work}
\label{sec:relatedwork}

Deception-driven defense has seen significant progress through various frameworks, each contributing valuable insights toward detecting and responding to cyber threats. Notable examples include H-DOCTOR~\cite{amal_h-doctor_2023}, SMASH~\cite{dambrosio_smash_2025}, and the distributed firewall tuning model by Tudosi et~al.~\cite{tudosi_design_2023}. While these frameworks introduce important innovations such as real-time detection and dynamic rule adaptation, they generally target on-premise or hybrid environments. Consequently, they fall short of fully meeting the requirements of cloud-native, automated defense pipelines, particularly in terms of scalability, integration with cloud-native services (e.g., Azure), and orchestration latency.

Other approaches, like HoneyIoT~\cite{guan_honeyiot_2023} and HoneyBoost~\cite{kandanaarachchi_honeyboost_2022}, emphasize deception fidelity and actively engaging attackers. However, these frameworks often stop short of transforming gathered intelligence into automated defense actions. Similarly, systems focused on cognitive deception, such as the Cyber Reconnaissance Deception System~\cite{achleitner_cyber_2016} and the Multi-Paradigm Deception System~\cite{de_faveri_multi-paradigm_2018}—increase attacker confusion but do not extend detection results into real-time firewall policy updates. Meanwhile, performance-oriented engines like the Rule Optimization Framework~\cite{trabelsi_dynamic_2014} enhance firewall efficiency but lack integration with deception-based threat intelligence.

This divide between detection and automated mitigation is a recurring theme in the literature. Many existing systems treat deception outputs as passive data streams rather than actionable triggers that can dynamically adjust defenses. Moreover, cloud-native automation capabilities, such as Azure Logic Apps, Sentinel Analytics, and REST API-driven firewall management—are rarely leveraged to their full potential. As a result, there is a clear need for frameworks that combine deception telemetry with adaptive firewall tuning to enable automated, real-time threat response in the cloud. 
In addition, frameworks such as HoneyBoost~\cite{kandanaarachchi_honeyboost_2022} and SMASH~\cite{dambrosio_smash_2025} use simulations to model attacker behavior and response strategies. Simulations provide repeatability and safety but may lack realism against sophisticated adversaries. 

%Dynamic rule-tuning frameworks, exemplified by H-DOCTOR~\cite{amal_h-doctor_2023}, employ containerized honeypots combined with fuzzy logic to create near real-time detection-to-response loops. While promising, these solutions often face scaling challenges within cloud-native ecosystems due to limited orchestration support. More recent methodologies have shifted toward cloud-integrated, serverless architectures that rely on API-driven automation and platform-native tools~\cite{tudosi_design_2023, reddy_future_2022}. Such approaches offer strong potential for enterprise scalability and reliability but require sophisticated integration to maintain policy consistency and operational trust across multiple cloud subscriptions and regions.

%Despite these advances, several gaps remain. Few frameworks have been evaluated under cloud-scale conditions or across multi-subscription cloud environments with complex governance and role-based access controls. Additionally, automated enforcement without proper validation can introduce harmful misconfigurations. Very few existing systems incorporate safeguards such as confidence thresholds, simulation of rule impacts, or rollback mechanisms—features that are critical to ensure automated responses do not inadvertently disrupt legitimate traffic~\cite{fatima_enhancing_2024, moric_advancing_2025}.

In summary, while there are considerable progress in both deception-based detection and automated firewall tuning, an integrated, cloud-native solution that seamlessly connects these components remains unclear. This research addresses this gap by designing a modular, Azure-native framework that transforms passive deception data into immediate, adaptive defense actions, supporting scalable, multi-subscription cloud environments and closing the loop between attacker detection and automated mitigation.

\section{Proposed Approach}
\label{sec:proposed_approach}

\begin{figure*}[!h]
\centering
{\scriptsize
\resizebox{0.85\textwidth}{!}{% reduced from 0.9 to 0.85 for EDAS gutter compliance
\begin{tikzpicture}[
    node distance=0.6cm,
    box/.style={rectangle, draw=black, text width=3.5cm, align=center, font=\scriptsize, rounded corners, fill=white},
    arrow/.style={thick, -{Latex[length=1.2mm]}},
    group/.style={rectangle, draw=black, dashed, rounded corners, inner sep=0.15cm, fill=#1!20},
    outerbox/.style={rectangle, draw=black, dashed, fill=gray!10, rounded corners, inner xsep=0.4cm, inner ysep=0.65cm},
    title/.style={font=\bfseries\scriptsize, align=center}
]

% Phase 1: Attacker Interaction (xshift left)
\node[box] (access) [xshift=-6cm] {Initial Access\\(SSH Port 22)};
\node[box, right=2cm of access] (redirect) {Port Redirected to 2222\\(Cowrie Honeypot)};

% Phase 2: Log Collection & Detection
\node[box, below=1cm of access] (syslog) {Syslog Logs\\(Honeypot Events)};
\node[box, right=2cm of syslog] (agent) {Azure Monitor Agent\\(Log Analytics Workspace)};
\node[box, right=2cm of agent] (sentinel) {Analytics Rules\\(Azure Sentinel)};

% Phase 3: MITRE ATT&CK & Validation
\node[box, below=1.3cm of syslog] (mitre) {MITRE ATT\string&CK Integration\\(Tagging \& Enrichment)};
\node[box, right=2cm of mitre] (validation) {System Validation \& Performance\\(MTTB, Attack Volume, Engagement Time)};

% Phase 4: Incident Response & Mitigation
\node[box, below=2cm of mitre, xshift=6cm] (logicapp) {Triggered Logic App\\(Parse \& Enrich IP)};
\node[box, left=1.5cm of logicapp] (nsg) {NSG / Azure Firewall\\(Update Rules)};
\node[box, right=1.5cm of logicapp] (soc) {Send Alert to SOC\\(Azure Monitor / Email / Webhook)};

% Phase 5: Continuous Monitoring & Improvement
\node[box, below=1cm of logicapp] (evaluate) {Evaluate Rule Effectiveness\\(False Positives, Response Time, MTTB)};
\node[box, left=1.5cm of evaluate] (viz) {Visualization \& Reporting};
\node[box, right=1.5cm of evaluate] (optimize) {Optimize Thresholds \& Logic\\(Feedback to Sentinel/Logic App)};

% Background groups
\begin{pgfonlayer}{background}
    \node[outerbox, fit=(access)(redirect)(syslog)(agent)(sentinel)(mitre)(validation)(logicapp)(nsg)(soc)(evaluate)(viz)(optimize)] (outer) {};

    \node[group=green, fit=(access)(redirect)] (g1) {};
    \node[title] at ($(g1.north west)!0.5!(g1.north east)+(0,0.2)$) {Attacker Interaction \& Honeypot Engagement};

    \node[group=blue, fit=(syslog)(agent)(sentinel)] (g2) {};
    \node[title] at ($(g2.north west)!0.5!(g2.north east)+(0,0.2)$) {Log Collection \& Detection};

    \node[group=orange, fit=(mitre)(validation)] (g3) {};
    \node[title] at ($(g3.north west)!0.5!(g3.north east)+(0,0.1)$) {MITRE ATT\string&CK Integration \& System Validation};

    \node[group=yellow, fit=(logicapp)(nsg)(soc)] (g4) {};
    \node[title] at ($(g4.north west)!0.5!(g4.north east)+(0,0.2)$) {Incident Response \& Mitigation};

    \node[group=purple, fit=(evaluate)(viz)(optimize)] (g5) {};
    \node[title] at ($(g5.north west)!0.5!(g5.north east)+(0,0.2)$) {Continuous Monitoring \& Improvement};
\end{pgfonlayer}

% Arrows
\draw[arrow] (access) -- (redirect);
\draw[arrow] (access) -- (syslog);
\draw[arrow] (redirect) -- (agent);
\draw[arrow] (redirect.east) -- ++(3.85cm,0) -- (sentinel.north);
\draw[arrow] (syslog) -- (mitre);
\draw[arrow] (mitre) -- (validation);
\draw[arrow] (validation) -- ++(0,-2.4cm);
\draw[arrow] (logicapp) -- (nsg);
\draw[arrow] (logicapp) -- (soc);
\draw[arrow] (logicapp) -- (evaluate);
\draw[arrow] (evaluate) -- (viz);
\draw[arrow] (evaluate) -- (optimize);

% Extra routing for arrows
\coordinate (intersect) at ($(syslog)!0.5!(mitre)$);
\draw[arrow] (sentinel.south) |- (intersect);

\coordinate (intersect_agent) at ([yshift=0.2cm]intersect);
\draw[arrow] (agent.south) |- (intersect_agent);

\end{tikzpicture}
}
}
\caption{A proposed framework showing the honeypot engagement, detection, MITRE integration, response, and continuous improvement}
\label{fig:honeypot-workflow}
\end{figure*}
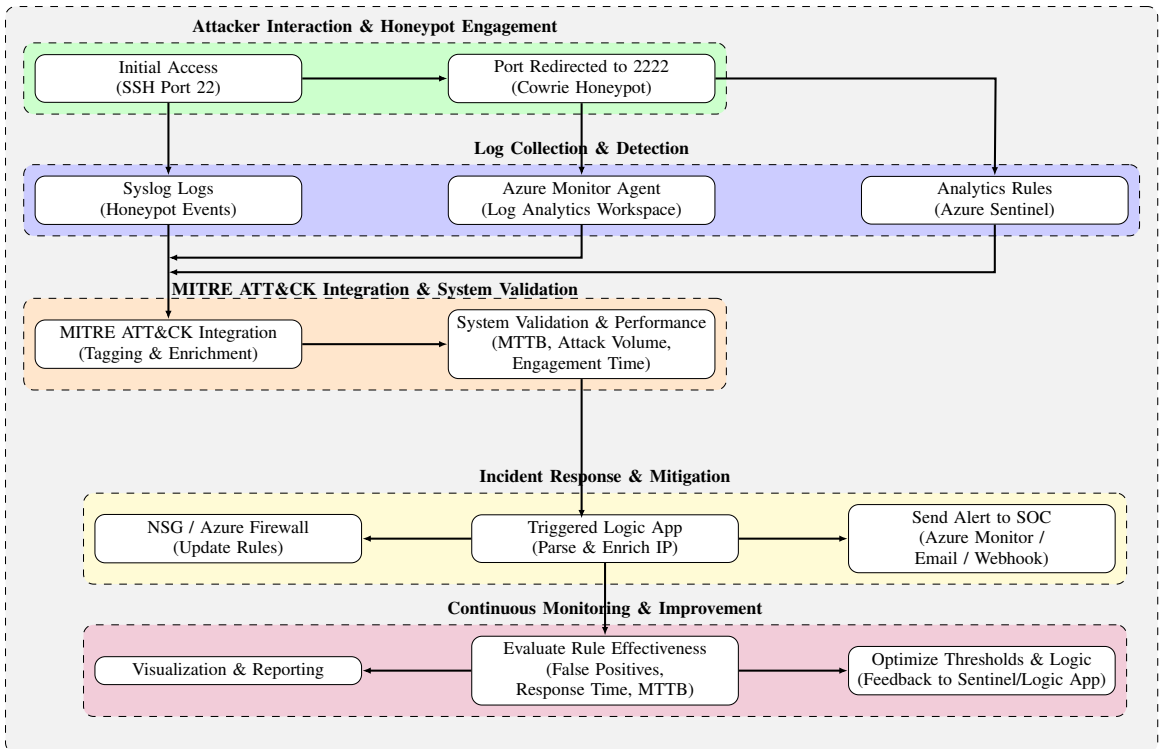

This section describes the proposed approach, demonstrating how attacker interactions captured by a medium- to high-interaction honeypot can drive dynamic policy enforcement using Azure-native tools. Figure~\ref{fig:honeypot-workflow} illustrates the framework and workflow for honeypot engagement and automated response.

The framework is divided into the following phases:

\subsection{Attacker Interaction \& Honeypot Engagement}
In the first phase, an Azure Linux VM is used to host a Cowrie honeypot within an isolated virtual network, where port 22 (SSH) is exposed externally to attract attackers and redirected internally to Cowrie on port 2222, while legitimate access is routed to a secured, obfuscated port (e.g., 8888). Cowrie emulates a real Ubuntu shell, logging usernames, passwords, commands, downloads, and session timings. Additional decoy services (e.g., Telnet) and anti-fingerprinting measures, such as modified SSH banners, further enhance engagement. All activity is captured via syslog and forwarded as telemetry, generating MITRE ATT\&CK-aligned data, including credential access attempts (T1110) and valid logins (T1078), which supports structured analysis in subsequent phases.

\subsection{Log Collection \& Detection}
Telemetry from the honeypot is forwarded into Azure-native monitoring services for near real-time analysis. This phase focuses on capturing and structuring data for subsequent processing:

\begin{enumerate}
    \item \textbf{System Logs:} The honeypot records endpoint activity, including authentication attempts, shell commands, file transfers, permission changes, and network scans.
    \item \textbf{Azure Monitor Agent (AMA):} The AMA securely forwards syslog data into an Azure Log Analytics workspace, where it is parsed into custom tables for structured correlation.
    \item \textbf{Analytics (Azure Sentinel):} Custom KQL rules detect suspicious activity. Key Cowrie events such as \texttt{cowrie.session.connect}, \texttt{cowrie.login.failed}, \texttt{cowrie.login.success}, \texttt{cowrie.session.closed}, \texttt{cowrie.client.version}, \texttt{cowrie.client.kex}, and \texttt{cowrie.command.input} are monitored for trend analysis and MITRE mapping in the next phase.
\end{enumerate}

\subsection{MITRE ATT\string&CK Integration \& System Validation}
This phase focuses on analyzing attacker behavior and validating system performance:

\begin{enumerate}
    \item \textbf{MITRE ATT\string&CK Integration:} Telemetry is mapped to MITRE ATT\string&CK tactics and techniques. Sentinel generates incidents containing source IPs, timestamps, command history, and triggering events, enabling automated response workflows. Table \ref{tab:mitre-mapping} shows the mapping of Cowrie events to MITRE ATT\string&CK techniques.

    \item \textbf{System Validation \& Performance Metrics:} Metrics such as Mean Time to Block (MTTB), attack volume, successful logins, and engagement duration are evaluated to ensure feasibility and responsiveness. Near real-time analytic rules reduce latency, enabling immediate log processing.
\end{enumerate}

\begin{table}[h]
\centering
\scriptsize
\caption{Cowrie Event-to-MITRE ATT\&CK Mapping}
\label{tab:mitre-mapping}
\resizebox{0.95\columnwidth}{!}{% reduced from \columnwidth for gutter safety
\begin{tabular}{|p{3.5cm}|p{2.0cm}|p{1.2cm}|p{3.0cm}|}
\hline
\textbf{EventID} & \textbf{Tactic} & \textbf{Tech ID} & \textbf{Technique} \\
\hline
cowrie.login.failure & Credential Access & T1110 & Brute Force \\
cowrie.login.success & Initial Access & T1078 & Valid Accounts \\
\multirow{4}{*}{cowrie.command.input} & Execution & T1059 & Command \& Scripting Interpreter \\
 & Discovery & T1082 & System Info Discovery \\
 & Discovery & T1083 & File \& Directory Discovery \\
 & Discovery & T1087 & Account Discovery \\
cowrie.client.version & Recon & T1046 & Network Service Scan \\
cowrie.client.kex & Recon & T1046 & Network Service Scan \\
\hline
\end{tabular}%
}
\end{table}

\subsection{Incident Response \& Mitigation}
Upon incident generation by Azure Sentinel, an automated workflow ensures rapid containment and network protection:

\begin{enumerate}
    \item \textbf{NSG/Azure Firewall:} The Logic App updates firewall or NSG rules via REST API, retrieving existing rules, adding malicious IPs, and applying updates for near real-time blocking.
    \item \textbf{Triggered Logic App:} The Logic App extracts relevant entities (e.g., source IPs), validates them, orchestrates firewall updates, and logs success or failure for traceability.
    \item \textbf{SOC Alerts:} Notifications are sent to the Security Operations Center via email, webhook, or Azure Monitor, keeping human operators informed. Continuous logging ensures auditability.
\end{enumerate}

This forms a closed-loop, adaptive response pipeline where deception data directly drives preventive action. Integration with Azure-native tools reduces manual overhead, scales across subscriptions, and supports rollback logic and threshold tuning.

\subsection{Security Analysis \& Continuous Improvement}
After initial response, the system enters continuous monitoring and optimisation to maintain agility and efficiency:

\begin{enumerate}
    \item \textbf{Evaluate Rule Effectiveness:} Firewall rules are analysed using Sentinel Workbooks and KQL to track blocked IP activity, engagement reduction, and mitigation success.
    \item \textbf{Visualization \& Reporting:} Dashboards show incident trends, automation success, top attacker sources, and MTTB improvements. Scheduled reports support audits, compliance, and executive visibility.
    \item \textbf{Optimize Thresholds \& Policy Refinement:} Response thresholds are adjusted to reduce false positives and unnecessary blocks, using strategies such as interaction thresholds, session duration filtering, geographic whitelisting, and rule expiry.
\end{enumerate}

This feedback loop ensures the defense pipeline dynamically adapts to the threat landscape, maximizing security effectiveness while maintaining operational resilience.

\section{Evaluation}
\label{sec:evaluation}
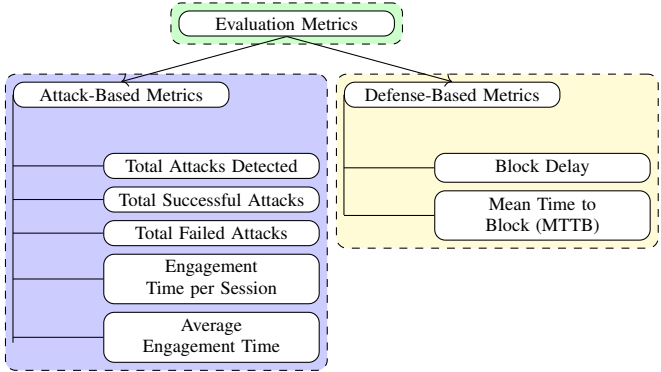
\begin{figure}[h!]
\centering
\scriptsize
\begin{tikzpicture}[
    node distance=0.3cm,
    box/.style={rectangle, draw=black, text width=2.7cm, align=center, font=\scriptsize, rounded corners, fill=white},
    group/.style={rectangle, draw=black, dashed, rounded corners, inner sep=0.1cm, fill=#1!20},
    title/.style={font=\bfseries\scriptsize, align=center}
]

% Level 1
\node[box] (eval) {Evaluation Metrics};

% Level 2 (separate level)
\node[box, below=0.6cm of eval, xshift=-2.2cm] (attack) {Attack-Based Metrics};
\node[box, below=0.6cm of eval, xshift=2.2cm] (defense) {Defense-Based Metrics};

% Level 3 nodes (right of vertical line)
\node[box, below=0.6cm of attack, xshift=1.2cm] (total_detected) {Total Attacks Detected};
\node[box, below=0.1cm of total_detected, xshift=0cm] (total_success) {Total Successful Attacks};
\node[box, below=0.1cm of total_success, xshift=0cm] (total_failed) {Total Failed Attacks};
\node[box, below=0.1cm of total_failed, xshift=0cm] (engage_time) {Engagement Time per Session};
\node[box, below=0.1cm of engage_time, xshift=0cm] (avg_engage) {Average Engagement Time};

\node[box, below=0.6cm of defense, xshift=1.2cm] (block_delay) {Block Delay};
\node[box, below=0.1cm of block_delay, xshift=0cm] (mttb) {Mean Time to Block (MTTB)};

% Connect Level 1 to Level 2
\draw[->] (eval.south) -- (attack.north);
\draw[->] (eval.south) -- (defense.north);

% Vertical lines from left of Level 2 to Level 3 nodes
\draw[-] (attack.west) -- ++(0,-0.1) coordinate (line_attack_start);
\draw[-] (line_attack_start) -- ++(0,-3.2cm) coordinate (line_attack_end); % extended further
\foreach \y in {total_detected,total_success,total_failed,engage_time,avg_engage}
    \draw[-] (line_attack_end |- \y.west) -- (\y.west);

\draw[-] (defense.west) -- ++(0,-0.1) coordinate (line_defense_start);
\draw[-] (line_defense_start) -- ++(0,-1.5cm) coordinate (line_defense_end);
\foreach \y in {block_delay,mttb}
    \draw[-] (line_defense_end |- \y.west) -- (\y.west);

% Groups
\begin{pgfonlayer}{background}
    \node[group=green, fit=(eval)] (g1) {};

    \node[group=blue, fit=(attack)(total_detected)(total_success)(total_failed)(engage_time)(avg_engage)] (g2) {};

    \node[group=yellow, fit=(defense)(block_delay)(mttb)] (g3) {};
\end{pgfonlayer}

\end{tikzpicture}
\caption{Evaluation Metrics - Categorisation of the Evaluation metrics used: Attack-Based and Defence-Based.}
\label{fig:model_evaluation}
\end{figure}

To evaluate the proposed approach from both attack and defense perspectives, we developed a set of security metrics, categorized into attack-based and defense-based metrics (Figure~\ref{fig:model_evaluation}), and applied them systematically during evaluation.

\subsection{Attack-Based Metrics}

To evaluate how effectively the honeypot captures adversary behavior, key attacker-centric metrics derived from Cowrie logs are defined. These metrics quantify attack attempts, success rates, and engagement times.

\subsubsection{Total Attacks Detected}
is the number of SSH connection attempts captured by the honeypot and it is given by equation~\eqref{eq:total_attacks}.

\begin{equation}
\text{Total Attacks} = \left| \text{cowrie.session.connect} \right|
\label{eq:total_attacks}
\end{equation}

\subsubsection{Total Successful Attacks}
is the number of times attackers successfully logged into the honeypot and it is given by equation~\eqref{eq:successful_attacks}.

\begin{equation}
\text{Successful Attacks} = \left| \text{cowrie.login.success} \right|
\label{eq:successful_attacks}
\end{equation}

\subsubsection{Total Failed Attacks}
is the number of failed login attempts to the honeypot and it is given by equation~\eqref{eq:failed_attacks}.
\begin{equation}
\text{Failed Attacks} = \left| \text{cowrie.login.failed} \right|
\label{eq:failed_attacks}
\end{equation}

\subsubsection{Attacker Engagement Time per Session}
is the duration an attacker remains connected during a single session ((from cowrie.session.closed)) and it is given by equation~\eqref{eq:engagement_time_session}.

\begin{equation}
\text{Engagement Time}_{\text{session } i} = \text{Duration}_i
\label{eq:engagement_time_session}
\end{equation}

\subsubsection{Average Attacker Engagement Time}
is the average session duration across all attacker sessions and it is given by equation~\eqref{eq:avg_engagement_time}.
\begin{equation}
\text{Average Engagement Time} = \frac{\sum_{i=1}^{n} \text{Duration}_i}{n}
\label{eq:avg_engagement_time}
\end{equation}

\subsection{Defence-Based Metrics}

These metrics measure how quickly the system responds once an attack is detected.

\subsubsection{Block Delay}
is the time between a successful attacker login and when their IP is blocked by the NSG via Logic App and it is given by equation~\eqref{eq:block_delay}.
\begin{equation}
\text{Block Delay}_i = T_{\text{NSG.Logic.App.Success},i} - T_{\text{NSG.Logic.App.Start},i}
\label{eq:block_delay}
\end{equation}

\subsubsection{Mean Time to Block (MTTB)}
is the average duration from attacker login to IP blocking and it is given by equation~\eqref{eq:mttb}.
\begin{equation}
\text{MTTB} = \frac{\sum_{i=1}^{n} \left( T_{\text{NSG.Logic.App.Success},i} - T_{\text{NSG.Logic.App.Start},i} \right)}{n}
\label{eq:mttb}
\end{equation}

\section{Experimental Results and Analysis}
\label{sec:Experiments}

This section presents results from a deception-driven, real-time firewall automation system implemented using Cowrie honeypot telemetry in Microsoft Azure. Data were collected over seven days (15–19 May and 22–23 May 2025), excluding maintenance downtime on 20–21 May. The analysis is divided into attack-based and defense-based metrics to illustrate attacker behavior and system responsiveness.

\subsection{Attack-Based Metrics}

\subsubsection{Total Attacks Detected}
\begin{figure}[ht]
    \centering
    \includegraphics[width=\columnwidth]{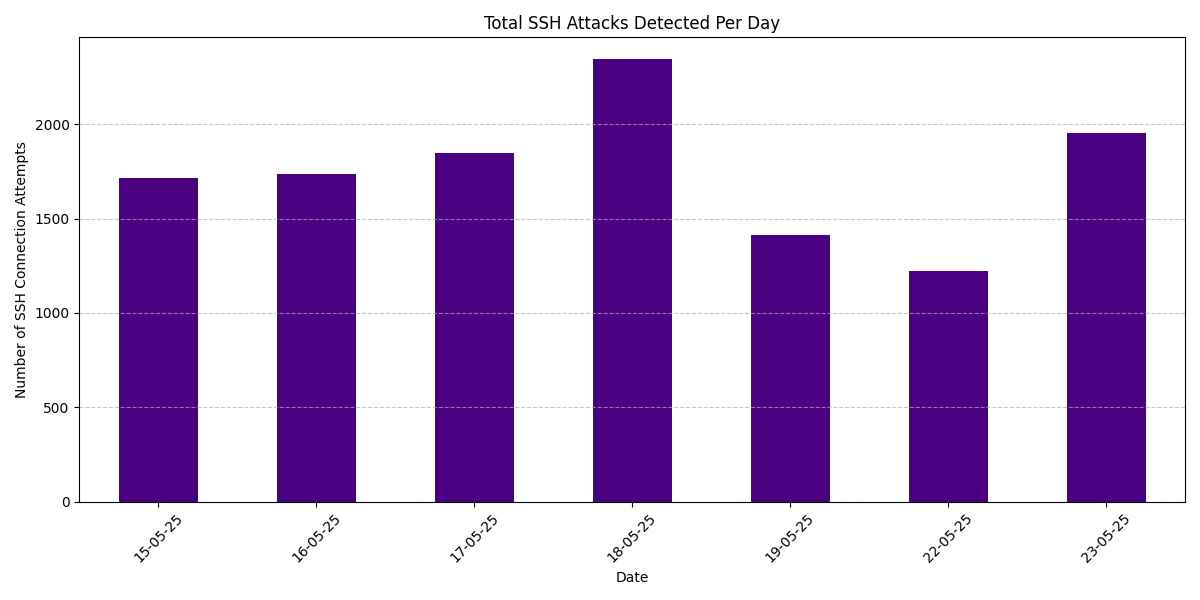}
    \caption{Total SSH Attacks Detected per Day — Peaks on 18 and 23 May 2025.}
    \label{fig:total_attacks}
\end{figure}

The honeypot captured a total of 12,224 SSH connection attempts over the observation period. Daily peaks occurred on 18 and 23 May, likely due to botnet activity or scheduled scanning campaigns (Figure~\ref{fig:total_attacks}). This high volume demonstrates the prevalence of automated reconnaissance and the importance of active monitoring.

\subsubsection{Successful vs Failed Logins}
\begin{figure}[ht]
    \centering
    \includegraphics[width=\columnwidth]{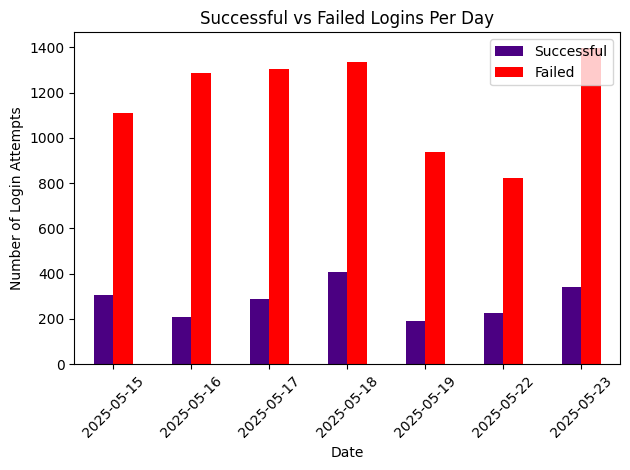}
    \caption{Successful vs Failed SSH Logins per Day — Red bars: failures, Indigo bars: successes.}
    \label{fig:s_vs_f_logins}
\end{figure}

Of the total connections, 2,008 were successful logins and 9,292 failed, yielding a 4.6:1 failure-to-success ratio. Figure~\ref{fig:s_vs_f_logins} illustrates daily trends, showing alignment with total attack spikes. This confirms widespread brute-force activity with occasional successful intrusions.

\subsubsection{Engagement Time per Session}
\begin{figure}[ht]
    \centering
    \includegraphics[width=\columnwidth]{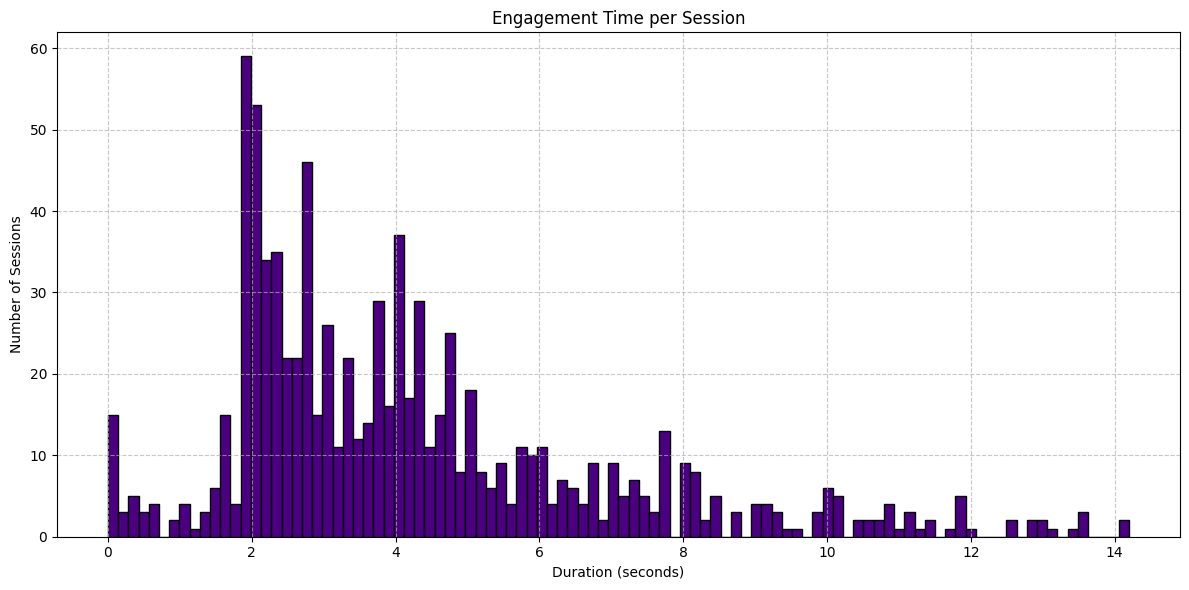}
    \caption{Histogram of Session Engagement Time — Most sessions under 5 seconds.}
    \label{fig:engagement_hist}
\end{figure}

\begin{figure}[ht]
    \centering
    \includegraphics[width=\columnwidth]{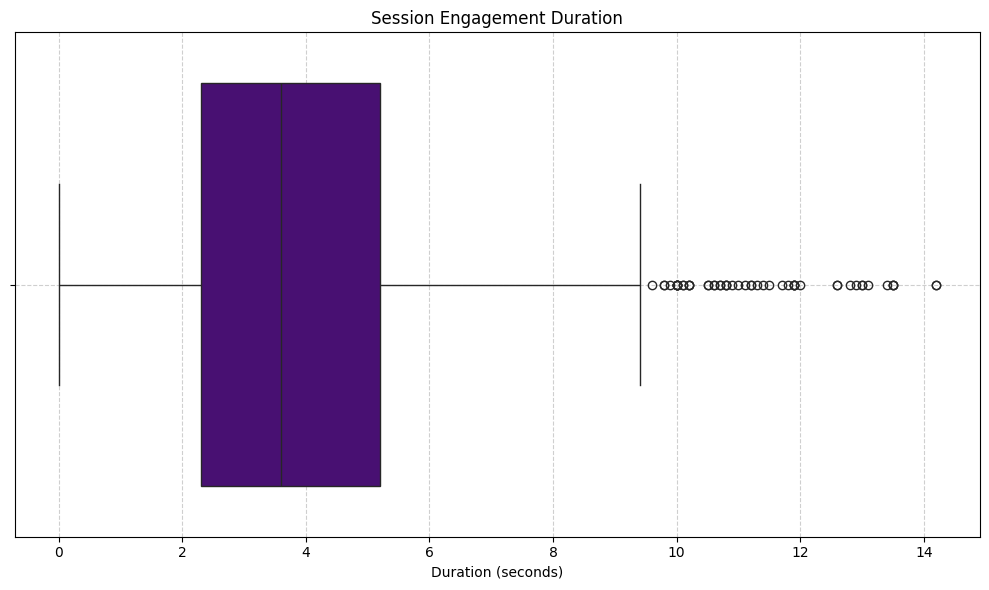}
    \caption{Box Plot of Engagement Duration — Clustered near median with long-tailed outliers.}
    \label{fig:engagement_box}
\end{figure}

Session duration provides insight into attacker intent and sophistication. After removing outliers above 9.5 seconds using the IQR method, the mean engagement time was 4.23 seconds (median 3.6 seconds), with most sessions under 5 seconds (Figures~\ref{fig:engagement_hist}, \ref{fig:engagement_box}). Short sessions suggest automated scripts, while long-tail outliers may indicate human-driven activity or misconfigured sessions.

\subsection{Defence-Based Metrics}

\subsubsection{Block Delay}
Block Delay measures the time from attacker login to IP blocking, and we calculate it by equation \eqref{eq:block_delay_}.

\begin{equation}
\text{Block Delay}_{i} = T_{\text{NSG.Logic.App.Success},i} - T_{\text{NSG.Logic.App.Start},i}
\label{eq:block_delay_}
\end{equation}

\begin{figure}[ht]
    \centering
    \includegraphics[width=\columnwidth]{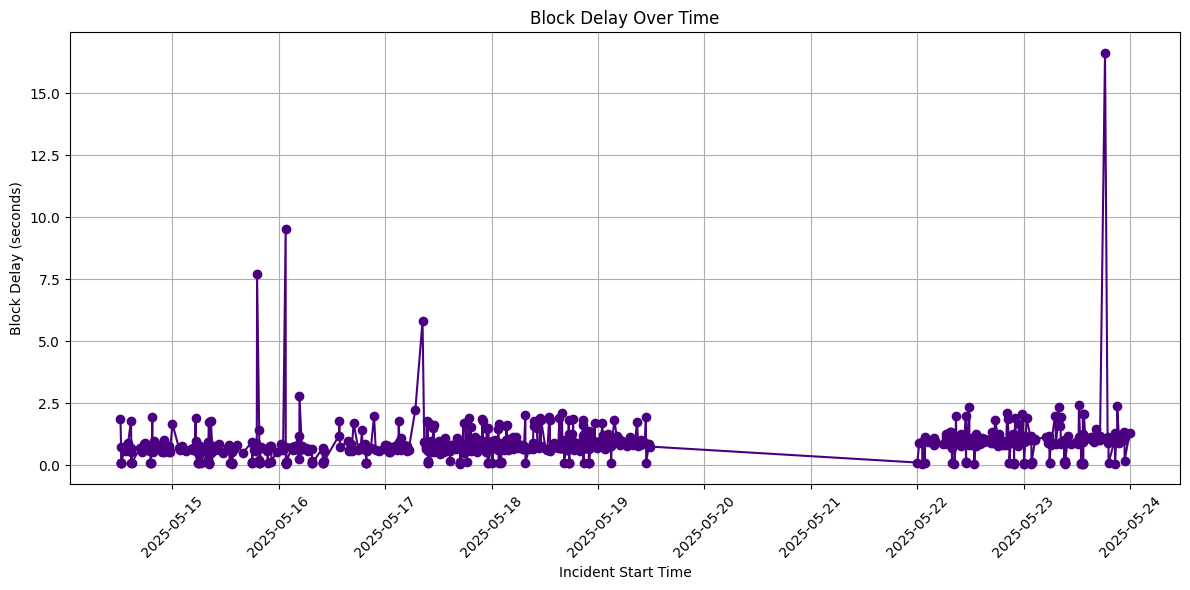}
    \caption{Block Delay Over Time — Most events under 2.5 seconds; outliers up to 16 seconds.}
    \label{fig:block_delay_time}
\end{figure}

\begin{figure}[ht]
    \centering
    \includegraphics[width=\columnwidth]{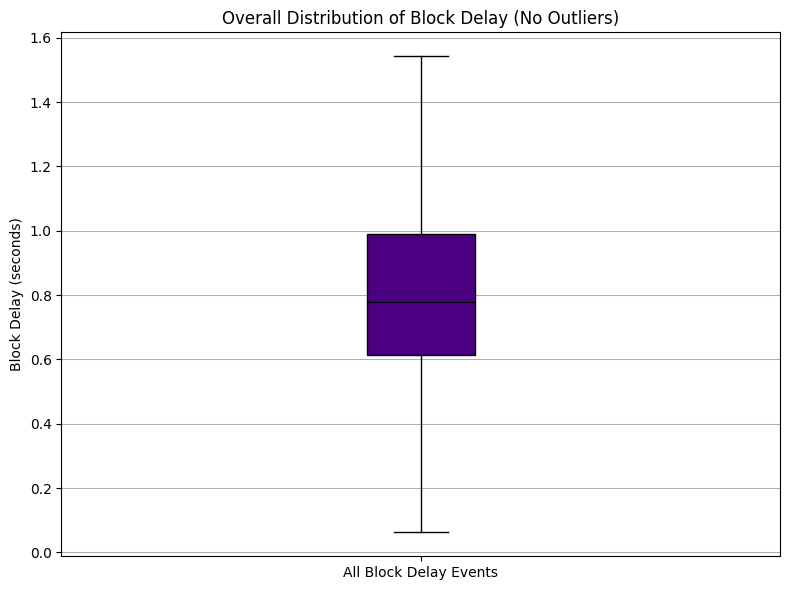}
    \caption{Box Plot of Block Delay Distribution — Median at 0.78 seconds.}
    \label{fig:block_delay_boxplot}
\end{figure}

The results show that the median Block Delay was 0.78 seconds, with most events under 2.5 seconds and occasional outliers up to 16 seconds (Figures~\ref{fig:block_delay_time}, \ref{fig:block_delay_boxplot}).

\subsubsection{Mean Time to Block (MTTB)}
Mean Time to Block represents average responsiveness to attacker detected, and we calculate it by equation \eqref{fig:mttb}.

\begin{equation}
\text{MTTB} = \frac{\sum_{i=1}^{n}(T_{\text{NSG.Logic.App.Success},i} - T_{\text{NSG.Logic.App.Start},i})}{n}
 \label{fig:mttb}
\end{equation}

The results show that the MTTB was \textbf{0.86 seconds}, indicating rapid detection and blocking of attackers.

\subsection{Summary}
Key findings include:
\begin{itemize}
    \item High-volume SSH probes (\textasciitilde12,224 over 7 days) with temporal spikes.
    \item Predominantly failed login attempts (9,292) vs successful logins (2,008), consistent with brute-force attacks.
    \item Short attacker dwell times (mean 4.23 seconds), highlighting automated tools, while long-tail sessions provide behavioral insights.
    \item Rapid containment via automated firewall orchestration, with median Block Delay of 0.78 seconds and MTTB of 0.86 seconds.
\end{itemize}

These results confirm that deception-driven telemetry combined with cloud-native automation enhances visibility and supports low-latency real-time mitigation.

\section{Attack Description and MITRE ATT\&CK Categorization}
\label{sec:MITRE}

\begin{figure}[h]
    \centering
    \includegraphics[width=\columnwidth]{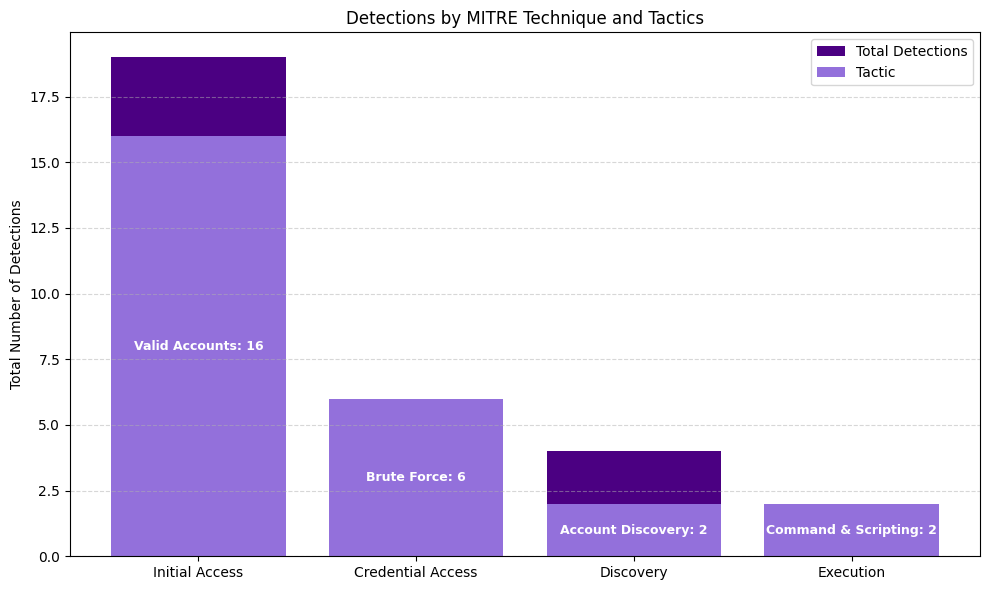}
    \caption{Total detections by MITRE tactic and technique.}
    \label{fig:TechniqueTacticsTotalDetection}
\end{figure}

To categorize the attacks, we developed custom scripts to map Cowrie telemetry data to MITRE ATT\&CK tactics and techniques. The key events captured are summarized in Table~\ref{tab:tactic_summary} and Figure~\ref{fig:TechniqueTacticsTotalDetection}.

%\subsection{Detection Summary}

Table~\ref{tab:tactic_summary} presents a summary of adversary activity, supporting SOC prioritization and enabling automated response. The table provides the total number of incidents observed for each tactic and technique. Across the dataset, the key observations are:

- **Initial Access**: 414 events, including 16 successful login events (\texttt{cowrie.login.success}).  
- **Credential Access**: 31 brute-force login attempts (\texttt{cowrie.login.failed}) mapped to T1110.  
- **Execution**: 68 command execution attempts (\texttt{cowrie.command.input}), mapped to T1059 and T1623.  
- **Discovery**: 451 system and network reconnaissance events (\texttt{cowrie.client.version}) mapped to T1046 and other discovery techniques.  

\begin{table}[h]
\centering
\scriptsize
\caption{Summary of tactic-level alert trends and incident counts.}
\begin{tabularx}{\columnwidth}{|l|c|X|}
\hline
\textbf{Tactic} & \textbf{Incidents} & \textbf{Description} \\
\hline
Initial Access & 414 & Automated scanning and valid credential logins (\texttt{cowrie.login.success}). \\
Credential Access & 31 & Brute-force logins mapped to T1110 (\texttt{cowrie.login.failed}). \\
Execution & 68 & Command execution attempts (\texttt{cowrie.command.input}), mapped to T1059 and T1623. \\
Discovery & 451 & System/network reconnaissance events mapped to T1046 and other discovery techniques. \\
\hline
\end{tabularx}
\label{tab:tactic_summary}
\end{table}

Figure~\ref{fig:TechniqueTacticsTotalDetection} shows the distribution of detections by MITRE tactic and technique. Initial Access and Discovery had the highest detection counts, followed by Execution and Credential Access. This pattern reflects realistic attacker engagement across multiple kill-chain stages.

\section{Conclusion}
\label{sec:conclusion}

We propose an automated defense framework leveraging medium–high interaction honeypot telemetry, Azure-native automation, and MITRE ATT\&CK-aligned detection. Cowrie sensors, Microsoft Sentinel, and Logic Apps enable real-time classification of attacker behavior and rapid network-level mitigation with minimal human intervention. Experimental evaluations demonstrate a sub-second MTTB (0.86 seconds), outperforming benchmark systems such as SMASH \cite{dambrosio_smash_2025}, H-DOCTOR \cite{amal_h-doctor_2023}, and traditional IDPS \cite{fatima_enhancing_2024}. Over 12,000 SSH attempts were mapped across MITRE ATT\&CK tactics, illustrating enhanced SOC visibility and actionable threat intelligence. The results highlight that combining honeypot telemetry with cloud-native automation reduces attacker dwell time while delivering scalable, real-time, and effective cloud security.

\addcontentsline{toc}{section}{References}

\sloppy
\bibliographystyle{IEEEtran}
\bibliography{mybibfile}

\end{document}